\documentclass[aps,prl,reprint,superscriptaddress,showpacs]{revtex4-1}  
\usepackage{booktabs}
\usepackage{graphicx}  
\usepackage{dcolumn}   
\usepackage{pbox}
\usepackage{bm}        
\usepackage{amssymb}   
\usepackage{amsmath}

\begin{document}
\renewcommand{\arraystretch}{1.5}
\mathchardef\mhyphen="2D

\title{An $^{27}$Al$^{+}$ quantum-logic clock with systematic uncertainty below $10^{-18}$}


\author{S.~M.~Brewer}
\email{samuel.brewer@nist.gov}
\affiliation{ Time and Frequency Division, National Institute of Standards and Technology, Boulder, CO 80305}
\affiliation{ Department of Physics, University of Colorado, Boulder, CO 80309}
\author{J.-S.~Chen}
\altaffiliation[Current Address: ]{IonQ, Inc., College Park, MD 20740}
\affiliation{ Time and Frequency Division, National Institute of Standards and Technology, Boulder, CO 80305}
\affiliation{ Department of Physics, University of Colorado, Boulder, CO 80309}
\author{A.~M.~Hankin}
\altaffiliation[Current Address: ]{Honeywell Quantum Solutions, Broomfield, CO 80021}
\affiliation{ Time and Frequency Division, National Institute of Standards and Technology, Boulder, CO 80305}
\affiliation{ Department of Physics, University of Colorado, Boulder, CO 80309}
\author{E.~R.~Clements}
\affiliation{ Time and Frequency Division, National Institute of Standards and Technology, Boulder, CO 80305}
\affiliation{ Department of Physics, University of Colorado, Boulder, CO 80309}
\author{C.~W.~Chou}
\affiliation{ Time and Frequency Division, National Institute of Standards and Technology, Boulder, CO 80305}
\author{D.~J.~Wineland}
\affiliation{ Time and Frequency Division, National Institute of Standards and Technology, Boulder, CO 80305}
\affiliation{ Department of Physics, University of Colorado, Boulder, CO 80309}
\affiliation{ Department of Physics, University of Oregon, Eugene, OR 97403}
\author{D.~B.~Hume}
\affiliation{ Time and Frequency Division, National Institute of Standards and Technology, Boulder, CO 80305}
\author{D.~R.~Leibrandt}
\email{david.leibrandt@nist.gov}
\affiliation{ Time and Frequency Division, National Institute of Standards and Technology, Boulder, CO 80305}
\affiliation{ Department of Physics, University of Colorado, Boulder, CO 80309}

\date{\today}

\begin{abstract}
We describe an optical atomic clock based on quantum-logic spectroscopy of the $^1$S$_0$ $\leftrightarrow$ $^3$P$_0$ transition in $^{27}$Al$^{+}$ with a systematic uncertainty of ${9.4 \times 10^{-19}}$ and a frequency stability of ${1.2\times10^{-15}/\sqrt{\tau}}$. A $^{25}$Mg$^{+}$ ion is simultaneously trapped with the $^{27}$Al$^{+}$ ion and used for sympathetic cooling and state readout. Improvements in a new trap have led to reduced secular motion heating, compared to previous $^{27}$Al$^{+}$ clocks, enabling clock operation with ion secular motion near the three-dimensional ground state. Operating the clock with a lower trap drive frequency has reduced excess micromotion compared to previous $^{27}$Al$^{+}$ clocks. Both of these improvements have led to a reduced time-dilation shift uncertainty. Other systematic uncertainties including those due to blackbody radiation and the second-order Zeeman effect have also been reduced.  
\end{abstract}


\maketitle

In 1973, Hans Dehmelt proposed a frequency standard based on a single trapped ion, dubbed the ``mono-ion oscillator", based on the $^{1}$S$_{0} \leftrightarrow ^{3}$P$_{0}$ transition in Tl$^{+}$ \cite{Dehmelt1973APSBulletin, Dehmelt1975APSBulletin}.  Sideband cooling was later added to this proposal \cite{Wineland1975APSBulletin} and, in 1982, the proposal was expanded to include B$^{+}$, Al$^{+}$, Ga$^{+}$ and In$^{+}$ \cite{Dehmelt1982IEEETransInstMeas.}.  In \cite{Dehmelt1982IEEETransInstMeas.} the possibility of a clock with a fractional frequency uncertainty of $10^{-18}$ was first discussed, setting the stage for a series of experiments that continue to push the limits of measurement science.  For trapped-ion systems, the systematic uncertainty was predicted to be limited by uncertainty in second-order Doppler (time-dilation) shifts due to the ion motion.

At this level of systematic uncertainty it is possible to measure clock frequency ratios that could lead to improved limits on the time-variation of fundamental constants, investigate dark matter composition, and probe physics beyond the standard model~\cite{Safronova2018BSM}.  Additionally, systematic uncertainty of $10^{-18}$ is one of the criteria in the roadmap for a possible redefinition of the SI second based on an optical frequency standard \cite{Riehle2018Metrologia}.  Furthermore, since the current techniques used for the characterization of the Earth's geoid are limited at a level corresponding to height differences of a few cm corresponding to gravitational redshifts of a few times $10^{-18}$ \cite{Denker2018JOG}, it is possible to use optical clocks at this level to improve knowledge of the geoid \cite{McGrew2018Nature}.

Since the original optical frequency standard proposals, significant experimental progress has been made in both systematic uncertainty and stability \cite{ChouAlAlcomparison, Nicholson2015NatComm, Ludlow2015RMP, Huntemann2016PRL, McGrew2018Nature, Schioppo2017NatPhot, Oelker2019arxiv}.  However, the systematic uncertainty of some of the highest performance trapped-ion clocks has been limited by Doppler shifts \cite{Rosenband2008Science, ChouAlAlcomparison, Huntemann2016PRL} that arise from ion trap imperfections that cause excess micromotion (EMM) and thermal (secular) motion.

Here, we report the systematic uncertainty evaluation of an optical atomic clock based on quantum-logic spectroscopy of $^{27}$Al$^{+}$ with a fractional frequency uncertainty of ${\Delta \nu / \nu = 9.4 \times 10^{-19}}$, which is the lowest systematic uncertainty reported for any clock to date.  This is achieved by operating the clock close to the three-dimensional (3D) motional ground state utilizing a new trap design that reduces secular motion heating and with lower trap drive and secular frequencies to reduce EMM compared to previous $^{27}$Al$^{+}$ systems, resulting in an order-of-magnitude reduction in uncertainty due to Doppler shifts \cite{Rosenband2008Science, ChouAlAlcomparison}.  In addition, we report a measurement of the clock stability, ${\sigma(\tau) = 1.2 \times 10^{-15}/\sqrt{\tau}}$.

\begin{figure}[]
\includegraphics[angle=0, width=\columnwidth]{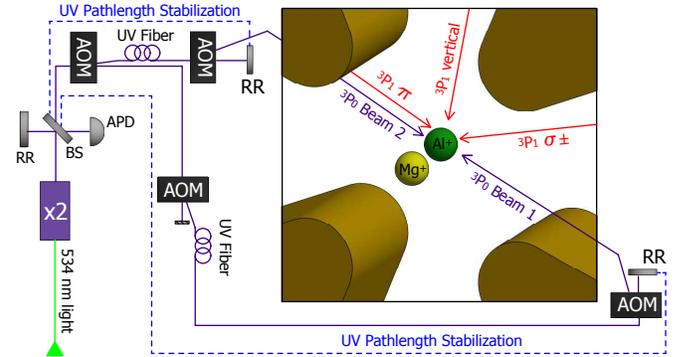}
\caption{\label{fig: expsetup} Simplified schematic of the quantum-logic clock experimental setup.  A frequency-quadrupled Yb-doped fiber laser is locked to the $^{1}$S$_{0} \leftrightarrow ^{3}$P$_{0}$ transition $(\lambda \simeq 267~\rm{nm})$ by alternating the probe direction between two counter-propagating laser beams (shown in violet).  An enlarged view of the trapping region is shown on the right.  Three nominally orthogonal beams used for micromotion measurements are shown in red.  Acousto-optic modulator (AOM); beam splitter (BS); retro-reflector (RR); frequency doubling stage~(x2).}
\end{figure}

\begin{figure}[]
\includegraphics[width=\columnwidth]{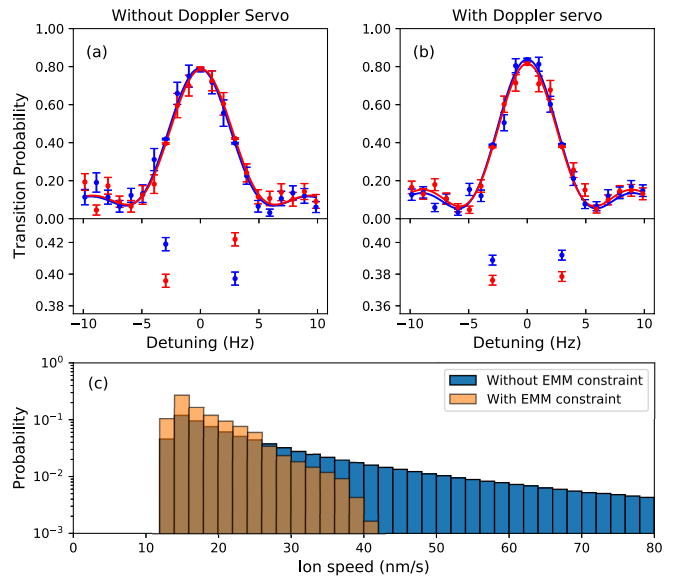}
\caption{\label{fig: 1stDopp} First-order Doppler shift characterization.  (a) Clock transition lineshapes for the two opposing probe directions (red and blue points) measured during clock operation without a first-order Doppler servo.  Solid lines show fits to a Rabi lineshape.  The zoomed in view shows that the transition probabilities at the probe frequencies used for the frequency lock are not balanced for each direction individually, indicating a non-zero first-order Doppler shift.  (b) Similar data taken while running the first-order Doppler servo, with balanced transition probabilities at the lock points.  (c) Distribution of possible ion speeds based on the measured first-order Doppler shift with and without an additional velocity constraint from EMM measurements.}
\end{figure}

The experimental setup, including the trap design and ground-state cooling (GSC) sequence, is described in detail elsewhere \cite{Chen2017PRL,Chen2017thesis,Supplement}.  A simplified schematic of the laser beams used to address $^{27}$Al$^{+}$ is shown in Fig.~\ref{fig: expsetup}.
The trap operates with a radiofrequency (RF) drive frequency of $\Omega_{RF}/2 \pi = 40.72$~MHz and a differential drive amplitude of approximately $\pm~30$~V.  The radial secular frequencies (motion perpendicular to the trap axis) of a single $^{25}$Mg$^{+}$ ion are ${\omega_{x}/2 \pi \approx 3.4}$~MHz and ${\omega_{y}/2 \pi \approx 4.0}$~MHz and the axial frequency is ${\omega_{z}/2 \pi \approx 1.5}$~MHz.  The clock operation sequence begins with preparation of the $^{27}$Al$^{+}$ state in either $|^1$S$_0$, $m_F = \pm5/2 \rangle$ by optical pumping on the $^1$S$_0 \leftrightarrow ^3$P$_1$ transition.  Next, the $^{25}$Mg$^{+}$/$^{27}$Al$^{+}$ pair is cooled to near the 3D motional ground state using 1~ms of far-detuned ($\Delta / 2\pi=-415$~MHz) laser cooling, 2~ms of Doppler cooling ($\Delta / 2\pi = -20$~MHz), and $\approx 12$ ms of pulsed Raman sideband cooling applied to the $^{25}$Mg$^{+}$ ion \cite{Chen2017PRL,Chen2017thesis,Supplement}.  Finally, a 150~ms clock interrogation pulse is applied to the $^{27}$Al$^{+}$ ion, followed by quantum-logic readout \cite{Schmidt2005Science,Hume2007}.  The clock is operated using Rabi spectroscopy with a Fourier-limited linewidth and $\approx 70~\%$ contrast (Fig.~\ref{fig: 1stDopp}).  

The $^{27}$Al$^{+}$ ion is interrogated alternately on the ${|^1\rm{S}_0, m_F = \pm 5/2 \rangle \leftrightarrow |^3\rm{P}_0, m_F = \pm 5/2 \rangle}$ transitions to generate a clock frequency that is to first-order insensitive to external magnetic fields \cite{Rosenband2007PRL}.  In addition to clock interrogation, auxiliary operations are interleaved to stabilize the orientation of the ion pair, track the $^{1}$S$_{0} \leftrightarrow ^{3}$P$_{1}$ frequency, and compensate excess micromotion (EMM) in real-time.  The clock duty cycle is $\approx 50~\%$, with $\approx 45~\%$ devoted to cooling, state preparation, and readout and $\approx 5~\%$ for auxiliary operations.

Systematic frequency shifts and associated uncertainties are listed in Table~\ref{tab: totunc}.  In previous $^{27}$Al$^{+}$ clocks, the dominant systematic uncertainty was due to EMM \cite{Rosenband2008Science,ChouAlAlcomparison}.  To evaluate the EMM shift and uncertainty we use the resolved-sideband technique \cite{Berkeland1998JAP, Keller2015JAP}.  The time-dilation shift $\Delta \nu/\nu$ due to EMM measured in a particular direction is given by
\begin{equation}
\label{eq: TDshift}
\frac{\Delta \nu}{\nu} = -\frac{\langle v^{2}_{EMM} \rangle}{2c^{2}} = -\left(\frac{\Omega_{RF}}{\omega_{L}}\right)^{2} \left(\frac{\Omega^{(\pm1)}_{EMM}}{\Omega^{(0)}} \right)^{2},
\end{equation}
where $v_{EMM}$ is the velocity of the ion in the direction of the probe beam k-vector, $c$ is the speed of light, ${\omega_{L} = 2 \pi c / \lambda_{L}}$ is the probe laser frequency, and ${\Omega^{(0)} (\Omega^{(\pm1)}_{EMM})}$ is the carrier (micromotion sideband) Rabi rate of the atomic transition.  In addition to the time-dilation shift, there exists an AC Stark shift due to the trap RF drive field.  The time-dilation shift and the RF drive AC Stark shift add to give the total frequency shift due to EMM \cite{ChouAlAlcomparison},
\begin{equation}
\label{eq: EMMfreqtot}
\frac{\Delta \nu}{\nu} = - \frac{\langle v^{2}_{EMM} \rangle}{2c^{2}} \left[1 + \left(\frac{\Omega_{RF} / 2 \pi}{400~\rm{MHz}} \right)^{2} \right] ,
\end{equation}
where the second term contributes approximately 1 \% to the total shift at $\Omega_{RF}/2\pi = 40.72$~MHz.

\begin{figure}[]
\includegraphics[width=\columnwidth]{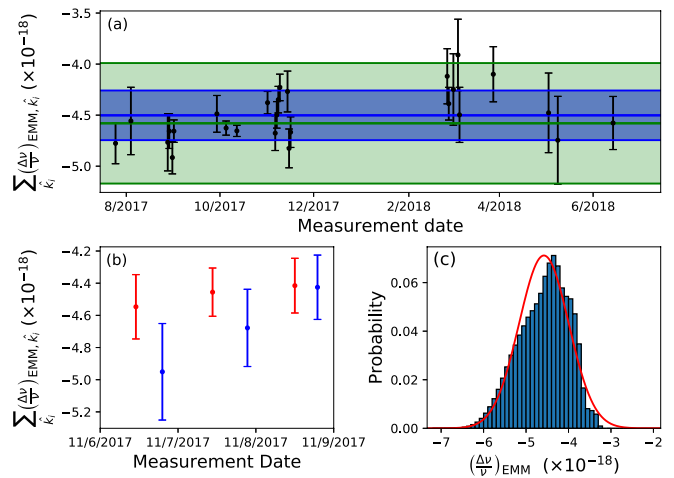}
\caption{\label{fig: EMMtot} Excess micromotion (EMM) shift evaluation.  (a) The sum of the EMM frequency shifts (black points) measured along three nearly orthogonal probe directions from August 2017 to June 2018, with the average and standard deviation (blue line and band). (b) Sample of EMM measurements on three consecutive days.  Red points are data taken immediately after initial EMM compensation and blue points are data taken after $\approx$ 12 hours of clock operation with interleaved micromotion compensation servos.  (c) Histogram of possible total EMM shifts consistent with the measurements generated by a Monte-Carlo analysis accounting for non-orthogonality of the probe directions and assuming the worst-case scenario of either 0 or $\pi$ phase between ion motion along these directions.  The total EMM shift of ${\Delta \nu / \nu = -(45.8 \pm 5.9) \times 10^{-19}}$ is given by the mean and standard deviation of the calculated distribution and shown in (a) (green line and band).  For reference, the red line shown in (c) is a normal (Gaussian) distribution with the same mean and standard deviation.}
\end{figure}

Measurements of the EMM were made on the $^{27}$Al$^{+}$ ion using the $^{1}$S$_{0} \leftrightarrow ^{3}$P$_{1}$ transition at $\lambda_{L} = 267$~nm, with three nearly orthogonal beams (see Fig. \ref{fig: expsetup}).  Figures~\ref{fig: EMMtot}(a) and \ref{fig: EMMtot}(b) show the sum of the EMM shifts measured along the three probe directions, $\hat{k_{i}}$, given by $\sum_{\hat{k_{i}}} \left(\Delta \nu / \nu\right)_{EMM, \hat{k_{i}}}$.  The EMM shift has been observed to be stable during clock operation both long term (Fig.~\ref{fig: EMMtot}(a)) and over the course of a day (Fig.~\ref{fig: EMMtot}(b)) when compensated in real-time.
Based on these measurements, a histogram of possible time-dilation shifts (Fig.~\ref{fig: EMMtot}(c)) has been generated using a Monte-Carlo approach, which accounts for non-orthogonality of the probe beams and includes the statistical spread in the EMM measurements, uncertainty in $\vec{k}$ of the $^{27}$Al$^{+}$ $^{3}$P$_{1}$ beams, and ambiguity in the relative phase of the EMM components \cite{Supplement}.  These results, combined with additional systematic uncertainties including the sampling of intrinsic micromotion \cite{Supplement}, indicate an averaged EMM-induced frequency shift of $\Delta \nu / \nu = -(45.8 \pm 5.9) \times 10^{-19}$.

To mitigate the first-order Doppler shift due to motion of the ion that is correlated with the interrogation cycle, the clock transition is alternately interrogated with two laser beams that are approximately counterpropagating.  Both beams are switched on during every probe cycle, with one of the beams detuned by 100 kHz from the transition so as to interact negligibly with the ion.  Under these conditions, we expect that any stray electric fields caused by photo-electrons generated by the clock laser light will be uncorrelated with the probe direction.  Charging of surfaces inside the vacuum chamber due to 280~nm cooling light applied before the clock interrogation can also lead to time-dependent stray electric fields which cause ion motion.  We observe an average first-order Doppler shift of $|\Delta \nu / \nu| = 4.6 \times 10^{-17}$, by comparing the center-frequency offset between the two opposing probe directions, as shown in Fig.~\ref{fig: 1stDopp}(a).  

For exactly counterpropagating beams and identical (but frequency shifted) lineshapes for the two probe directions, the first-order Doppler shift does not shift the clock frequency.  If the spectroscopy lineshapes are different due to unequal intensity or phase noise on the two beams, the gain of the clock servo error signal will be different for the two probe directions.  For servo algorithms in which the two directions are probed with the same laser frequency, as used in previous $^{27}$Al$^{+}$ clocks and shown in Fig.~\ref{fig: 1stDopp}(a), this causes the output of the servo to be pulled closer to the probe direction that has higher contrast.  To eliminate this as a potential source of systematic uncertainty, we use a clock servo algorithm in which the resonance frequencies of the two probe directions are tracked independently, and the servo synthesizes the mean of these frequencies as its output, shown in Fig.~\ref{fig: 1stDopp}(b).  We have verified numerically that the servo error of our first-order Doppler tracking servo is much less than the statistical clock instability for all measurement times $>100$~s.

For perfectly counterpropagating probe beams, ion motion in any direction is exactly cancelled and does not contribute a systematic shift to the clock frequency.  However, in the case of misalignment of the two beams, the Doppler shift due to motion along the bisector of their k-vectors is not suppressed.  The two counterpropagating beams originate from UV fibers and are mode-matched on either side of the vacuum chamber to give approximately 60~\% transmission through each opposing fiber.  This contrains the angle between the wavefronts of the two clock beams at the location of the ion to be $\le 3$~mrad.  We impose a bound on the maximum possible ion velocity that is consistent with EMM measurements made of the ion displacement at various times during the clock interrogation sequence (Fig.~\ref{fig: 1stDopp}(c)).  From the average radial mode frequency and the EMM amplitude we deduce the average ion displacement away from the fully compensated location and corresponding speed \cite{Supplement}.  Based on this velocity constraint, we assign a first-order Doppler shift and uncertainty of $\Delta \nu / \nu = (0.0 \pm 2.2) \times 10^{-19}$.

The clock is operated with a bias magnetic field ${B \approx 0.12~\rm{mT}}$.  The quadratic Zeeman shift is given by ${\Delta \nu / \nu = C_{2} \langle B^{2} \rangle}$, where $C_{2}$ is the quadratic Zeeman coefficient and ${\langle B^{2} \rangle = \langle B_{DC}\rangle^{2} + \langle B_{AC}^{2}\rangle}$ \cite{Rosenband2008Science, ChouAlAlcomparison}.  Here $B_{DC}$ is the static magnetic field measured in real-time and $B_{AC}$ is constrained based on microwave frequency measurements made on the $^{25}$Mg$^{+}$ ion as well as the uncertainty in the hyperfine constant $A_{hfs}$ \cite{Gan2018PRA}.  We have recently made improved measurements of both $C_{2}$ and $A_{hfs}$ that are presented elsewhere \cite{Brewer2019PRA}.  The mean quadratic Zeeman shift for a day of operation is ${\Delta \nu / \nu = -(9241.8 \pm 3.7) \times 10^{-19}}$, where the exact value of the shift depends on the measured $B_{DC}$, but the uncertainty is not affected at the stated level of precision.

To reduce the frequency shift and uncertainty due to secular motion, the clock is operated close to the 3D motional ground state \cite{Chen2017PRL, Chen2017thesis}.  The sideband cooling sequence is chosen to ensure at least $90~\%$ of the remaining kinetic energy after Doppler cooling is removed \cite{Chen2017thesis}.  The characterization of the energy after sideband cooling is accomplished by comparing a numerical simulation of the cooling dynamics with experimental measurements of the ion temperature \cite{Chen2017PRL}.  The average occupation numbers of each motional mode estimated in \cite{Supplement} are used to calculate the time-dilation shift due to secular motion.  At a clock interrogation time $t_i$, the fractional time-dilation shift due to secular motion is
\begin{equation}
\frac{\Delta \nu}{\nu} = \sum_{p} \left(\frac{\Delta\nu_p}{\nu}\right)\left[\left(\frac{1}{2} + \bar{n}_{p,0}\right)+\frac{1}{2}\dot{\bar{n}}_p\,t_i\right]\text{,}
\end{equation}
where $(\Delta\nu_p/\nu)$ is the fractional time-dilation shift per quantum of motion in a particular secular mode $p$, and $\bar{n}_{p,0}$ and $\dot{\bar{n}}_p$ are the average occupation number after cooling and the heating rate, respectively.  The heating rate of each mode is measured using sideband thermometry \cite{Diedrich1989PRL, Monroe1995PRL} and the results are summarized in \cite{Supplement}.  For $150$~ms clock interrogation time, the time-dilation shift due to secular motion is $\Delta \nu / \nu = -(17.3 \pm 2.9) \times 10^{-19}$.

The $^{27}$Al$^{+}$ clock is operated in an apparatus held near room temperature ($\approx 295$~K) and the presence of blackbody radiation (BBR) leads to an AC Stark shift on the clock transition.  The clock frequency shift due to BBR depends on the sensitivity of the transition to thermal radiation, determined largely by the static differential polarizability, $\Delta \alpha_{clock}(0) = (7.02 \pm 0.95) \times 10^{-42} \rm{J}\rm{m^{2}}/\rm{V^{2}}$, and the temperature of the BBR at the position of the ion, $T_{BBR}$ \cite{Supplement, Safronova2011PRL}.  For an uncertainty in $T_{\rm{BBR}}$ below 9~K, the uncertainty in $\Delta \alpha_{clock}(0)$ is the dominant uncertainty in the BBR shift evaluation.  The temperature environment is characterized using seven thermocouple sensors; three located on the trap wafer and support structure and four located on the surrounding vacuum chamber \cite{Supplement}.  These measurements constrain the temperature at the ion to be $T_{\rm{BBR}} = (294.8 \pm 2.7)$~K.  The corresponding BBR induced frequency shift is evaluated as $\Delta \nu / \nu = -(30.5 \pm 4.2) \times 10^{-19}$.

\begin{table}[b]
\caption{\label{tab: totunc}
Fractional frequency shifts $(\Delta \nu / \nu)$ and associated systematic uncertainties for the $^{27}$Al$^{+}$ quantum-logic clock.}
\begin{ruledtabular}
\begin{tabular}{lcc}
Effect & Shift $(10^{-19})$ &    Uncertainty $(10^{-19})$ \\
\hline
Excess micromotion & -45.8 & 5.9 \\ [-0.75ex]
Blackbody radiation & -30.5 & 4.2 \\ [-0.75ex]
Quadratic Zeeman & -9241.8 & 3.7 \\ [-0.75ex]
Secular motion & -17.3 & 2.9 \\ [-0.75ex]
Background gas collisions & -0.6 & 2.4 \\ [-0.75ex]
First-order Doppler & 0 & 2.2 \\ [-0.75ex]
Clock laser Stark & 0 & 2.0 \\ [-0.75ex]
AOM phase chirp & 0 & $< 1$ \\ [-0.75ex]
Electric quadrupole & 0 & $< 1$ \\ [-0.75ex]
\hline
Total & -9336.0 & 9.4 \\
\end{tabular}
\end{ruledtabular}
\end{table}

Collisions of the $^{25}$Mg$^{+}$/$^{27}$Al$^{+}$ ion pair with background gas molecules cause both clock phase shifts and secular motion heating.  Here, we summarize the background gas collision shift and uncertainty; details are presented in \cite{Hankin2019BGC}.  We measure the pressure of H$_2$ background gas at the position of the ions to be $(3.8 \pm 1.9) \times 10^{-8}$~Pa by monitoring the rate of collisions that cause the two ions to swap positions.  Collisions of H$_2$ with either ion excite the secular motion into a non-thermal distribution with a tail extending out to near room temperature.  This high energy tail is too small to detect with sideband thermometry heating rate measurements, but Monte-Carlo simulations of the clock interrogation indicate that it contributes a time-dilation shift ${\Delta \nu / \nu = -0.6(^{+0.6}_{-0.3}) \times 10^{-19}}$ which for bookkeeping purposes we do not include in our secular motion shift.  When H$_2$ collides with $^{27}$Al$^{+}$ during the Rabi interrogation, the phase of the $^{27}$Al$^{+}$ superposition state is shifted, resulting in a spectroscopic frequency shift.  Since the magnitude of this phase shift is unknown for H$_2 / ^{27}$Al$^{+}$ collisions, we bound the collisional frequency shift by assuming the worst case $\pm \pi/2$ phase shift for Langevin spiraling collisions that penetrate the angular momentum barrier.  In this way, we constrain the collision shift to be $\Delta \nu / \nu = -(0.6 \pm 2.4) \times 10^{-19}$.

A possible AC stark shift due to the clock probe beams has previously been investigated \cite{ChouAlAlcomparison} and for the operating conditions used here, this leads to a clock laser induced AC stark shift of $\Delta \nu / \nu = (0.0 \pm 2.0) \times 10^{-19}$.  Other possible frequency shifts include those due to a phase chirp in the clock beam AOMs and an electric quadrupole shift due to the (static) axial trapping potential.  Uncertainties due to these shifts have been bounded below $10^{-19}$~\cite{Beloy2017PRA}.  

\begin{figure}[]
\includegraphics[angle=0, width=\columnwidth]{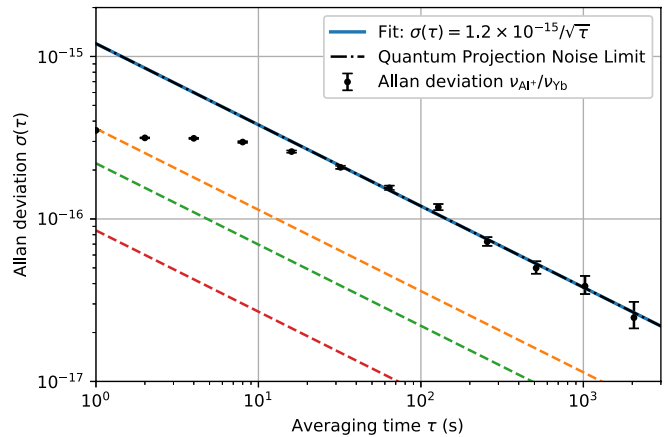}
\caption{\label{fig: AlYbStability} Allan deviation of the frequency ratio $\nu_{\rm{Al^{+}}} / \nu_{\rm{Yb}}$ measured over $\approx 23,000$~s.  The asymptote is fit to extract a frequency stability of $\sigma (\tau) = 1.2 \times 10^{-15} / \sqrt{\tau}$, where $\tau$ is the averaging time in seconds.  Anticipated stabilities for a correlation spectroscopy comparison of two single-ion $^{27}$Al$^{+}$ clocks \cite{Chou2011PRL} (orange), a single $^{27}$Al$^{+}$ mixed-species correlation comparison \cite{Hume2016PRA} (green), and a single $^{27}$Al$^{+}$ ion clock operated using Rabi spectroscopy at the interrogation time of 20.6 s, equal to the excited state lifetime (red) are also shown.}
\end{figure}

The $^{27}$Al$^{+}$ clock stability, measured by comparing with a Yb lattice clock at NIST, is shown in Fig.~\ref{fig: AlYbStability}.  The Yb clock has a stability of $\sigma (\tau) = 1.4 \times 10^{-16} / \sqrt{\tau}$; therefore, a measurement of the $\nu_{\rm {Al^{+}}} / \nu_{\rm{Yb}}$ frequency ratio provides a direct measure of the $^{27}$Al$^{+}$ clock stability \cite{McGrew2018Nature, Schioppo2016NaturePhot}.  The $^{27}$Al$^{+}$ clock beam pathlengths are stabilized from the output of the UV frequency doubler to the vacuum chamber \cite{Ma1994OL} and the probe time of 150 ms is chosen to optimize the stability.
The asymptotic stability is fit to $\sigma (\tau) = 1.2 \times 10^{-15} / \sqrt{\tau}$, consistent with the expected quantum projection noise \cite{Itano1993PRA}.  In the future, it should be possible to increase the probe time to achieve a single ion clock stability near $10^{-16} / \sqrt{\tau}$ with the use of a more stable clock laser \cite{Matei2018PRL}.

In conclusion, we have developed an $^{27}$Al$^{+}$ quantum-logic clock with a total systematic uncertainty of ${\Delta \nu / \nu = 9.4 \times 10^{-19}}$, fulfilling the vision of Dehmelt that a ``mono-ion oscillator" achieve a systematic uncertainty of $10^{-18}$.  The systematic uncertainty is limited by the uncertainty in the time-dilation shift due to excess micromotion.  Further improvements in trap design, uncertainty in the static differential polarizability, and a reduction in background gas pressure may lead to an improvement in the systematic uncertainty of the clock.  

We thank T.~Rosenband for development of the initial version of the ion trap used for this clock and useful discussions.  We thank K.~Beloy and J.~Bergquist for useful discussions, A.~Ludlow, W.~McGrew, and X.~Zhang for operating the Yb lattice clock, S.~Diddams, T.~Fortier, and H.~Leopardi for frequency measurements, and J.~Bollinger and C.~Oates for their careful reading of the manuscript.  This work was supported by the National Institute of Standards and Technology, the Defense Advanced Research Projects Agency, and the Office of Naval Research.  S.M.B. was supported by the U.S. Army Research Office through MURI Grant No. W911NF-11-1-0400.  This Letter is a contribution of the U.S. Government, not subject to U.S. copyright.

\bibliography{mainreferences}
\end{document}


\beginsupplement
\raggedbottom
\renewcommand{\arraystretch}{1.5}
\mathchardef\mhyphen="2D

\title{\textit{Supplemental Material for} \\ An $^{27}$Al$^{+}$ quantum-logic clock with systematic uncertainty below $10^{-18}$}

\author{S.~M.~Brewer}
\email{samuel.brewer@nist.gov}
\affiliation{ Time and Frequency Division, National Institute of Standards and Technology, Boulder, CO 80305}
\affiliation{ Department of Physics, University of Colorado, Boulder, CO 80309}
\author{J.-S.~Chen}
\altaffiliation[Current Address: ]{IonQ, Inc., College Park, MD 20740}
\affiliation{ Time and Frequency Division, National Institute of Standards and Technology, Boulder, CO 80305}
\affiliation{ Department of Physics, University of Colorado, Boulder, CO 80309}
\author{A.~M.~Hankin}
\altaffiliation[Current Address: ]{Honeywell Quantum Solutions, Broomfield, CO 80021}
\affiliation{ Time and Frequency Division, National Institute of Standards and Technology, Boulder, CO 80305}
\affiliation{ Department of Physics, University of Colorado, Boulder, CO 80309}
\author{E.~R.~Clements}
\affiliation{ Time and Frequency Division, National Institute of Standards and Technology, Boulder, CO 80305}
\affiliation{ Department of Physics, University of Colorado, Boulder, CO 80309}
\author{C.~W.~Chou}
\affiliation{ Time and Frequency Division, National Institute of Standards and Technology, Boulder, CO 80305}
\author{D.~J.~Wineland}
\affiliation{ Time and Frequency Division, National Institute of Standards and Technology, Boulder, CO 80305}
\affiliation{ Department of Physics, University of Colorado, Boulder, CO 80309}
\affiliation{ Department of Physics, University of Oregon, Eugene, OR 97403}
\author{D.~B.~Hume}
\affiliation{ Time and Frequency Division, National Institute of Standards and Technology, Boulder, CO 80305}
\author{D.~R.~Leibrandt}
\email{david.leibrandt@nist.gov}
\affiliation{ Time and Frequency Division, National Institute of Standards and Technology, Boulder, CO 80305}
\affiliation{ Department of Physics, University of Colorado, Boulder, CO 80309}

\date{\today}

\begin{abstract}
\end{abstract}

\maketitle

\section{Ion trap and laser beam geometry}
\begin{figure}[h]
\includegraphics[width=\columnwidth]{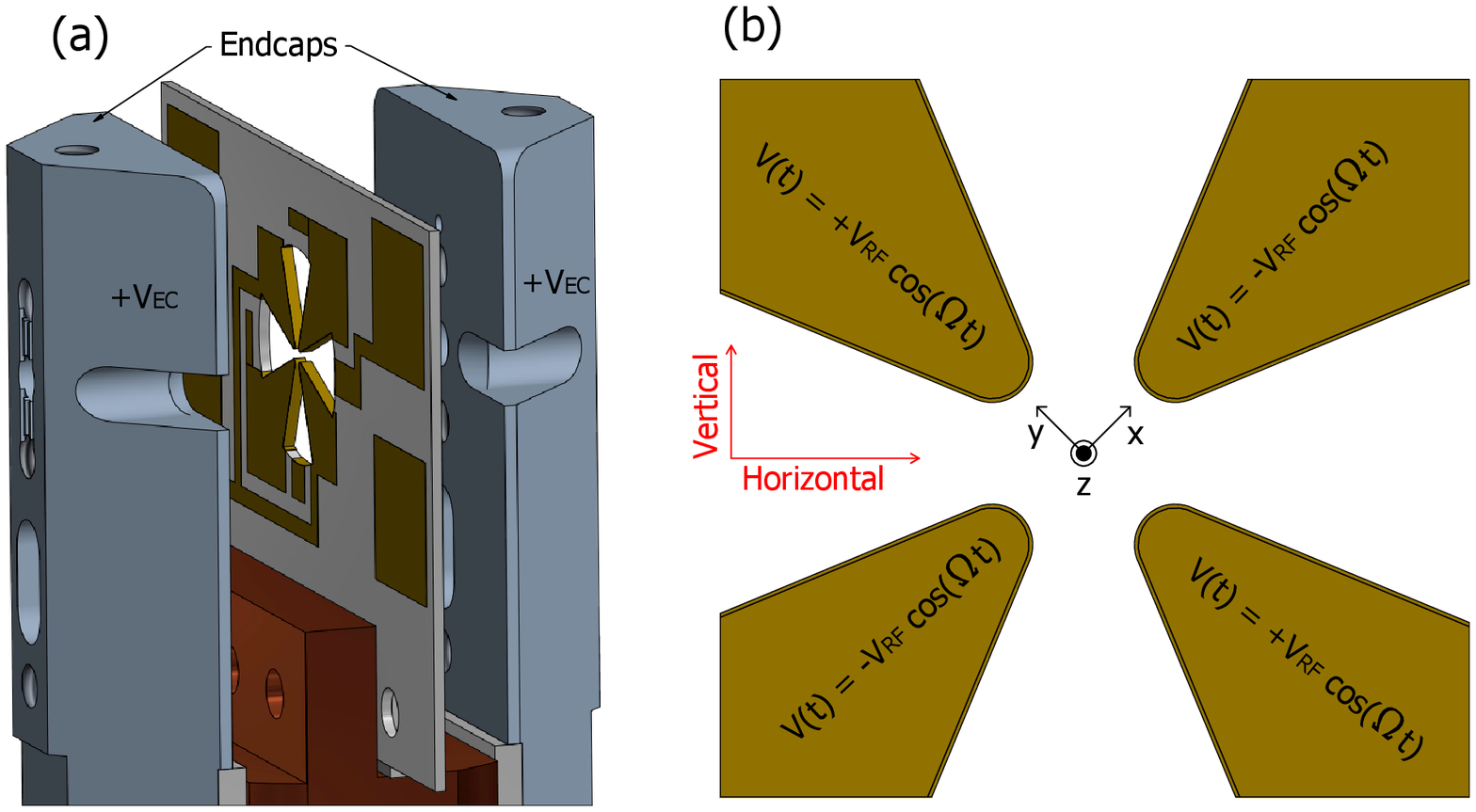}
\caption{\label{fig:trapscheme} Simplified schematic of the ion trap wafer and endcaps.  (a) The ion trap wafer and polished titanium endcaps are shown on the copper post used for mounting the trap structure in the vacuum chamber. A thermocouple sensor (not shown) is located on the bottom of the trap wafer and two additional thermocouples are mounted to the top and bottom of the trap mounting post.  (b) Zoomed in view of the trap wafer.  The normal mode coordinates $(x,y,z)$ are indicated (black) as well as the vertical and horizontal directions in the trap radial plane (red).}
\end{figure}

\begin{figure*}[]
\includegraphics[width=1.3\columnwidth, angle=90]{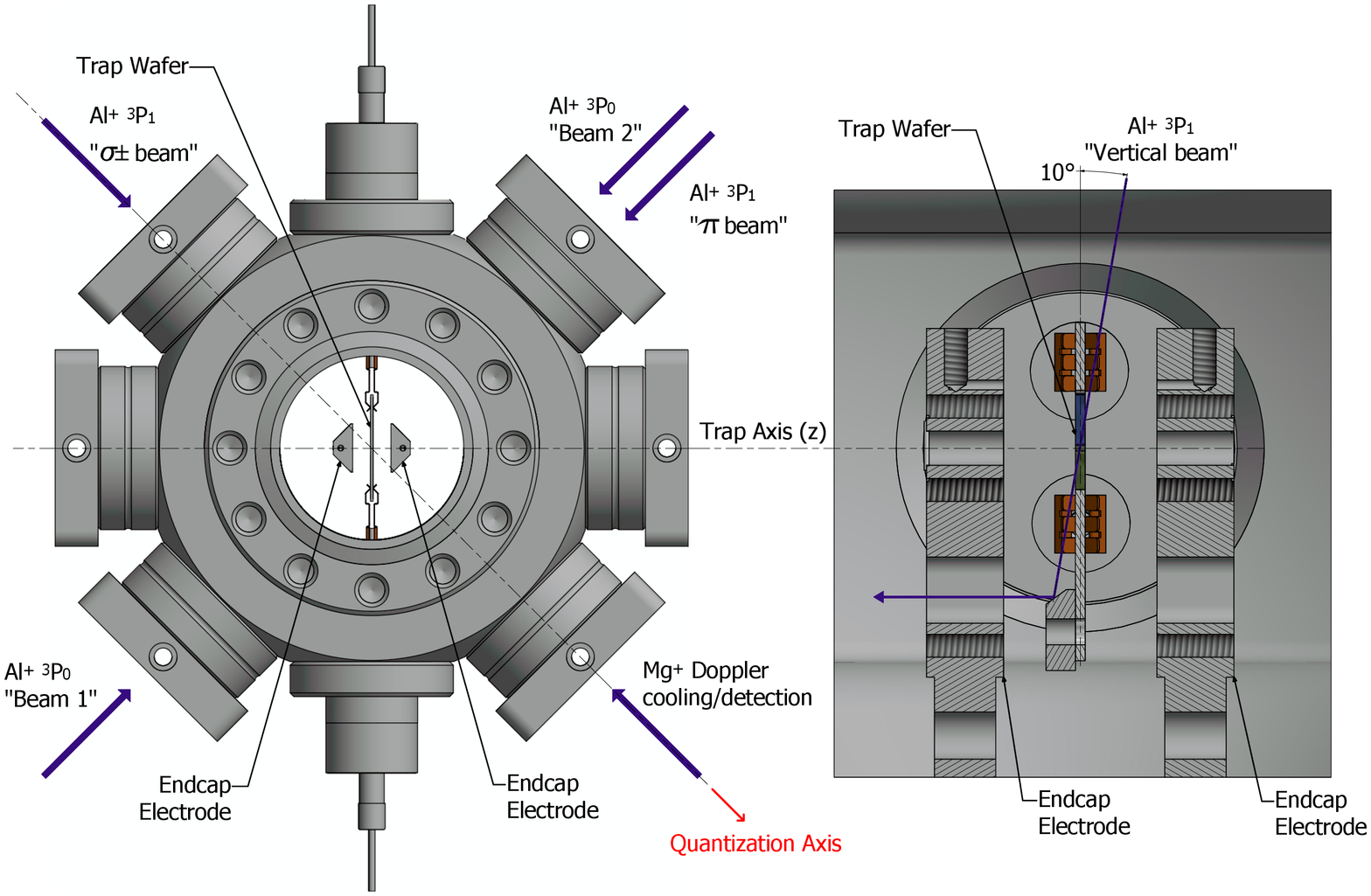}
\caption{\label{fig: Albeams} Trap geometry and laser beam layout.  (Left) Top-down view of the Al$^{+}$ clock ion trap and vacuum chamber.  The trap axis indicated here corresponds to the normal mode coordinate $z$ in Fig.~\ref{fig:trapscheme}.  The Al$^{+}$ $^{3}P_{1}$ beams used for state preparation, quantum-logic readout, and EMM characterization in the horizontal plane are shown.  The counterpropagating $^{3}P_{0}$ clock probe beams are also shown.  The $\approx 0.12$~mT quantization axis (red arrow) is generated by a pair of coils wrapped on the vacuum view port flanges.  Additional shim coils are located on the orthogonal view port flanges and are used to null any stray magnetic fields at the location of the ion.  (Right) Enlarged side section-view of the ion trap wafer and endcaps.  The Al$^{+}$ $^{3}P_{1}$ vertical beam used for EMM characterization is shown.  The beam has a nominal 10$^{\circ}$ angle with respect to the trap wafer, or 80$^{\circ}$ with respect to the trap axis.}
\end{figure*}

The ion trap and laser beam geometry are described in detail in \cite{Chen2017Thesis}.  Here, we discuss details relevant to the evaluation of the systematic uncertainties associated with the first-order Doppler shift, time-dilation shift due to secular motion and excess micromotion (EMM), and the blackbody radiation (BBR) shift.  A simplified schematic of the trap wafer and endcaps is shown in Fig.~\ref{fig:trapscheme}.
The trap wafer is constructed from a laser-machined $300~\mu\rm{m}$ thick diamond wafer that has been sputtered with gold to generate the radiofrequency (RF) trap electrodes and DC compensation electrodes.  As indicated in Fig.~\ref{fig:trapscheme}, the RF drive is applied differentially to generate the radial trapping potential and the endcap electrodes are held at positive DC potential relative to the trap wafer.  The trap wafer is mounted to a copper post, which is attached to a copper heat sink that is exposed to the ambient laboratory environment.  A simplified schematic of the trap wafer, vacuum chamber, and laser beam geometry is shown in Fig.~\ref{fig: Albeams}.\\

\newpage
\section{First-order Doppler Shift}
\subsection{Uncertainty due to error in beam alignment}
If the propagation directions of the two clock probe laser beams are not parallel, then the two beams will not see equal and opposite first-order Doppler shifts, and their average frequency will include a residual first-order Doppler shift.  We define a coordinate system shown in Fig.~\ref{fig: Albeams} such that the propagation direction of the first beam is given by
\begin{equation}
\hat{\kappa}_{1} = \hat{z'}
\end{equation}
and the propagation direction of the second beam is given by
\begin{equation}
\hat{\kappa}_{2} = -\sin(\alpha) \hat{x'} - \cos(\alpha) \hat{z'} \ .
\end{equation}
We assume that the ion velocity during the probe is equally likely to be in any direction, parameterized by
\begin{equation}
\vec{v} = v \sin(\theta) \cos(\phi) \hat{x'} +  v \sin(\theta) \sin(\phi) \hat{y'} + v \cos(\theta) \hat{z'} \ ,
\end{equation}
where $\theta \in [0,\pi]$ and $\phi \in [0,2\pi]$.  The first-order Doppler shift seen by beam $i$ is
\begin{equation}
\frac{\Delta \nu_i}{\nu} = -\frac{\hat{\kappa}_i \cdot \vec{v}}{c} \ .
\end{equation}
The measured first-order Doppler shift was
\begin{equation}
\frac{\Delta \nu_\textrm{measured}}{\nu} = \frac{1}{2} \left( \frac{\Delta \nu_1}{\nu} - \frac{\Delta \nu_2}{\nu} \right) = 4.6 \times 10^{-17} \ ,
\end{equation}
corresponding to a 13.8~nm/s velocity along the probe direction.  The residual uncancelled first-order Doppler shift as a function of the beam misalignment $\alpha$ and the velocity direction defined by $\theta$ and $\phi$ is given by
\begin{widetext}
\begin{align}
\frac{\Delta \nu_\textrm{residual}}{\nu} = \frac{1}{2} \left( \frac{\Delta \nu_1}{\nu} + \frac{\Delta \nu_2}{\nu} \right) = \frac{\cos(\theta) - \sin(\alpha) \sin(\theta) \cos(\phi) - \cos(\alpha) \cos(\theta)}{\cos(\theta) + \sin(\alpha) \sin(\theta) \cos(\phi) + \cos(\alpha) \cos(\theta)} \frac{\Delta \nu_\textrm{measured}}{\nu} \ .
\end{align}
\end{widetext}
Due to the divergence of this quantity when the velocity bisects the two laser beam directions, the standard deviation of the residual uncancelled Doppler shift taken over all possible ion velocity directions diverges for $\alpha \neq 0$.
To bound the ion velocity along the bisector, we use a Monte-Carlo scheme to generate distributions of possible ion velocities that are consistent with both the measured first-order Doppler shift and additional excess micromotion (EMM) measurements discussed below.  We use the 95 \% confidence interval of the distributions to derive the final uncertainty in the first-order Doppler shift.  The distributions of possible ion velocities are shown in Fig. 3 of the main text.

\subsection{Excess micromotion measurements of ion motion during the clock probe}
\begin{figure}[]
\includegraphics[width=1.0\columnwidth, angle=0]{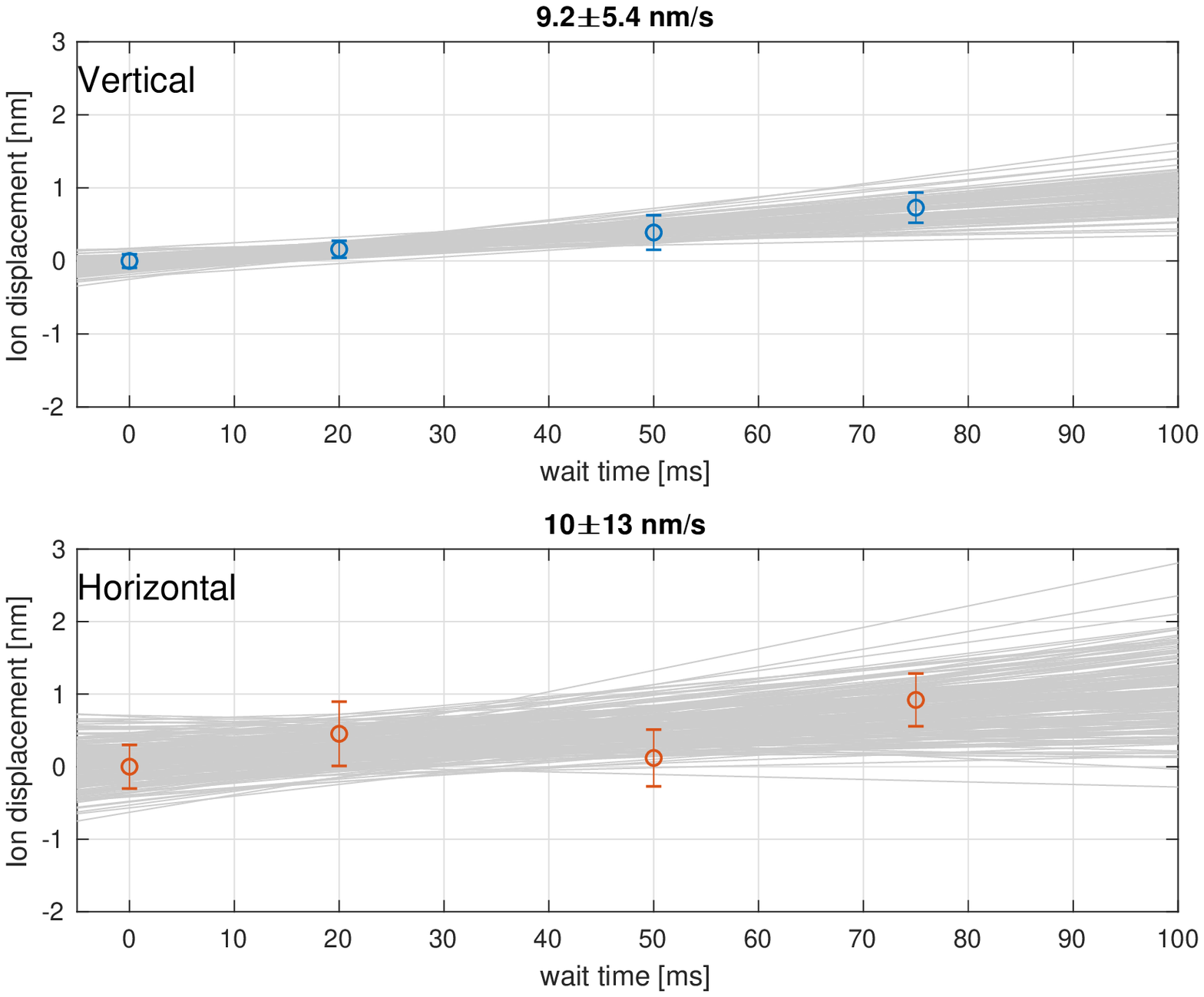}
\caption{\label{fig: EMMfirstorderDoppler}
Measurements of ion motion after initial GSC for EMM-based constraint on the radial velocity of the ion.  EMM measurements in the vertical direction (blue points) are shown in the upper panel and measurements in the horizontal direction (red points) are shown in the lower panel where the uncertainties in the points are statistical.  Gaussian resampling (grey lines) is used to estimate the velocity bounds from the measurements.  The labels at the top of each panel indicate the average speed and and 95 \% confidence interval for 1000 fit trials.}
\end{figure}

Measurements of the excess micromotion (EMM) were made as a function of the time after ground-state cooling (GSC) and the results are shown in Fig.~\ref{fig: EMMfirstorderDoppler}.  The cooling sequence, clock duty cycle, and readout sequence were all set to be identical to the conditions during typical clock operation.  From the amplitude of the observed EMM compared to the initially compensated values, we calculate the corresponding displacement of the ions.  Gaussian resampling of the ion displacement at each time is used to constrain the possible velocity of the ions orthogonal to the $^{27}$Al$^{+}$ $^{3}P_{0}$ beams.  From these measurements, we place a bound on the possible radial velocities of 16~nm/s in the vertical direction and 25~nm/s in the horizontal direction.

\subsection{Summary of the first-order Doppler shift evaluation}
From the average measured first-order Doppler shift, the constraint of the clock laser beam overlap, and the EMM constrained distribution of possible ion velocities during the clock probe, we evaluate a first-order Doppler shift and uncertainty of $\Delta \nu / \nu = (0.0 \pm 2.2) \times 10^{-19}$.

\section{Excess micromotion}
\subsection{Measurement}
The amplitude of the excess micromotion (EMM) is measured using the resolved-sideband technique \cite{Berkeland1998JAP, Keller2015JAP, Kellerthesis2015}.  Here we present a brief summary of this technique and measurements that have been made on the $^{27}$Al$^{+}$ ion.  When the ion is exposed to the radiofrequency (RF) driving field from the trap electrodes (i.e. either due to a stray field that pushes the ion away from the RF minimum or a phase imbalance that leads to a residual RF field at the RF minimum), the ion undergoes excess micromotion at the trap drive frequency $\Omega_{RF}$.  When the moving ion interacts with a probe laser, the ion experiences a modulated laser phase, leading to sidebands at the trap drive frequency with the amplitude of the laser field $E(\omega)$ to first order given by \cite{Berkeland1998JAP, Keller2015JAP, Kellerthesis2015}
\begin{multline}
\label{eq: atomspec}
E(\omega) \approx J_{0}(\beta)\delta(\omega - \omega_{L}) + J_{1}(\beta)[\delta(\omega - \omega_{L} + \Omega_{RF}) + \\
\delta(\omega - \omega_{L} - \Omega_{RF})],
\end{multline}
where $\delta$ is the Dirac delta function, $\omega_{L}$ is the probe laser frequency, and $J_{i}(\beta)$ is a Bessel function of the first kind corresponding to the carrier ($i = 0$) and first micromotion sideband ($i = 1$).  The modulation index is given by $\beta = (\vec{k}_{L} \cdot \vec{v}_{EMM})/\Omega_{RF}$, where $|\vec{k}_{L}| = 2\pi / \lambda_{L}$ and $\lambda_{L}$ is the probe laser wavelength.  For the case of low modulation index ($\beta \ll 1$), the ratio of the first micromotion sideband Rabi rate $\Omega^{(\pm1)}_{EMM}$ to the carrier Rabi rate $\Omega^{(0)}$ is given by
\begin{equation}
\label{eq: Rabiratio}
\frac{\Omega^{(\pm1)}_{EMM}}{\Omega^{(0)}} \approx \frac{\beta}{2}.
\end{equation}

\begin{figure}[]
\includegraphics[width=\columnwidth]{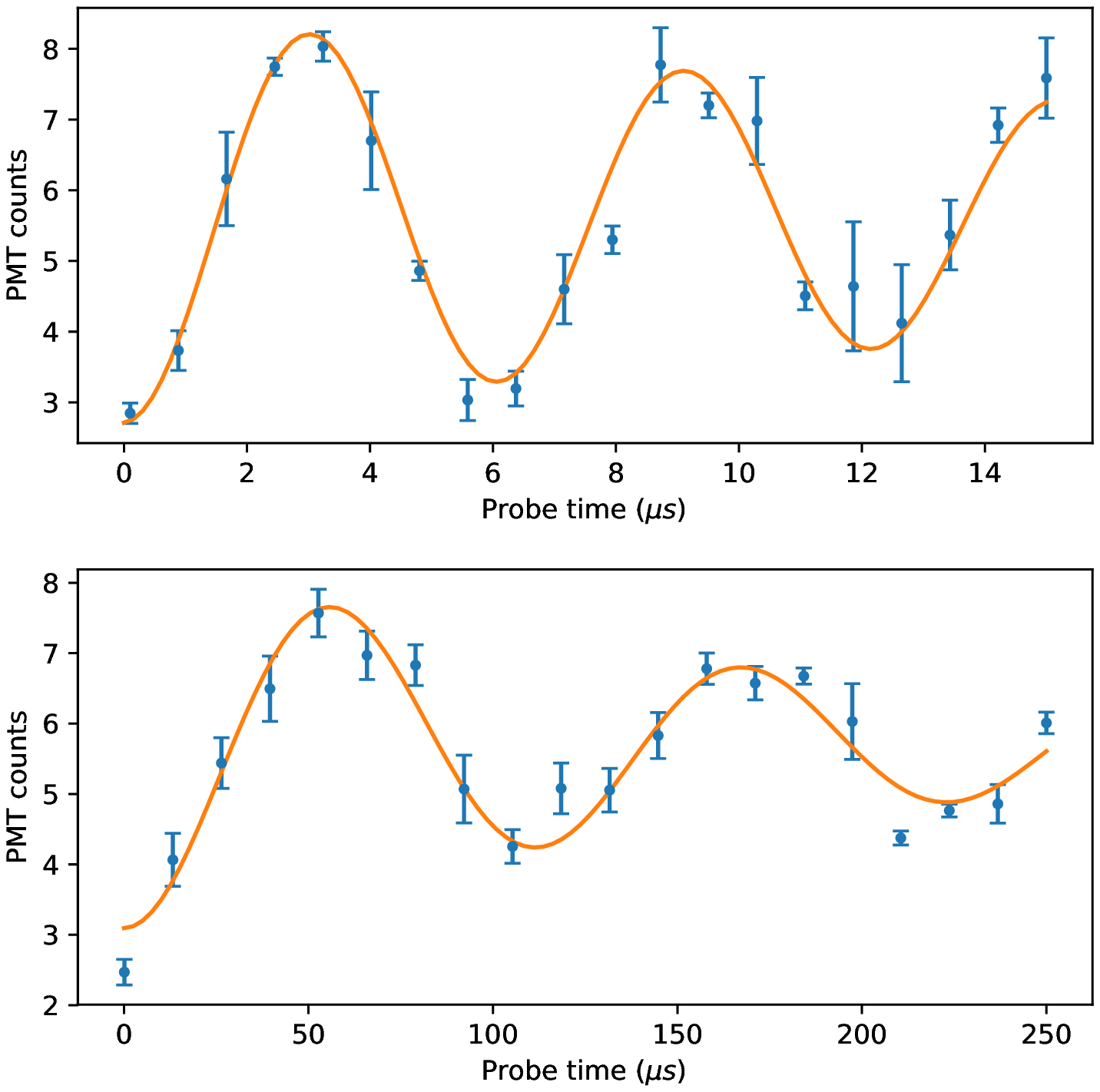}
\caption{\label{fig: EMMvert} Rabi flopping on the carrier (top panel) and micromotion sideband (bottom panel) of the $^{27}$Al$^{+}$ $^{1}S_{0} \leftrightarrow ^{3}P_{1}$ transition driven by the vertical beam.  The number of counts measured on a photomultiplier tube (PMT) in 125~$\mu$s of $^{25}$Mg$^{+}$ detection is roughly proportional to the $^{27}$Al$^{+}$ transition probability.  For this data, the frequency shift due to time-dilation is $\Delta \nu / \nu = -3.8 \pm 0.2 \times 10^{-18}$, where the uncertainty is given by the statistical uncertainties in the carrier and micromotion sideband fits.}
\end{figure}

An example of a micromotion measurement is shown in Fig.~\ref{fig: EMMvert}.
Here the measurement is performed using the vertical $^{3}P_{1}$ beam. (See Fig.~\ref{fig: Albeams})  There is a nominal 10$^{\circ}$ angle of the beam with respect to the trap wafer along the $z$ axis.  In order to improve the coherence of the transitions and ensure that the measured EMM is the same as that during the clock probe, the measurements are made immediately after three-dimensional (3D) GSC of the $^{27}$Al$^{+}$/$^{25}$Mg$^{+}$ pair.

\subsection{Systematic uncertainty in EMM measurements}

\subsubsection{Numerical estimates of micromotion and time-dilation uncertainty due to nonorthogonal beams}
An uncertainty in the k-vector directions of the three $^{27}$Al$^{+}$ $^{3}P_{1}$ beams coupled with an uncertainty in the relative phase of the EMM measured in three non-orthogonal directions leads to additional uncertainty in the total EMM time-dilation shift.  
We have carried out a Monte-Carlo simulation where we randomize the relative angles of the micromotion probe beams as well as the phases of the micromotion in each direction.  For the $^{3}P_{1}$ $\sigma \pm$ beam and $\pi$ beam (See Fig.~\ref{fig: Albeams}), we constrain the k-vector direction to within $\pm~5^{\circ}$ both vertically and horizontally, while the $^{3}P_{1}$ vertical-port beam nominally makes an 80$^{\circ}$ angle with the trap axis and is constrained geometrically to be within $\pm~2^{\circ}$.

To generate an upper bound uncertainty, we choose the relative phase of the measured electric fields to be either 0 or $\pi$.  The components of that field along the three measurement directions are
\begin{eqnarray}
E_1 = \vec{k}_1\cdot\vec{E}_0,\\
E_2 = \pm \vec{k}_2\cdot\vec{E}_0,\\
E_3 = \pm \vec{k}_3\cdot\vec{E}_0.
\end{eqnarray}
Defining an array of k-vectors,
\begin{equation}
\boldsymbol{k} \equiv \begin{pmatrix} \vec{k}_1 \\ \vec{k}_2 \\ \vec{k}_3\end{pmatrix},
\end{equation}
we solve for the unknown electric field amplitude and direction using,
\begin{equation}
\vec{E}_0 = \boldsymbol{k}^{-1}\begin{pmatrix} E_1 \\ \pm E_2 \\ \pm E_3\end{pmatrix}
\end{equation}

The distributions of possible micromotion amplitudes based on this Monte-Carlo treatment of the angle and phase uncertainties are shown in Fig. 2 of the main text.  This numerical treatment combines uncertainties associated with beam direction, motional phase and statistical fluctuations in the calibrated value.  The distributions produced by this analysis are somewhat asymmetric, primarily due to the nominal 10$^{\circ}$ projection of the $^3P_1$ vertical beam onto the trap axis.  This non-orthogonality, coupled with uncertainty in the relative phase of the micromotion measured in three directions, contributes the largest uncertainty in our analysis.  The total width of the distribution shown in Fig. 3 of the main text is $5.8\times10^{-19}$, which we take as the combined uncertainty due to uncertainty in the relative phase, k-vectors, and statistical spread of the measurements.


\subsubsection{Sampling of the intrinsic micromotion}
In an RF trap, the ion's intrinsic micromotion (IMM) is a result of the ponderomotive potential used to confine the ion.  
This leads to a non-vanishing IMM contribution to the measured EMM amplitude \cite{Keller2015JAP, Kellerthesis2015}.  When driving the micromotion sideband transition, the measured Rabi rate is given by
\begin{equation}
\label{eq: Omegatot}
| \Omega ^{(\pm 1)} |^{2} = | \Omega^{(\pm 1)}_{IMM} |^{2} + | \Omega^{(\pm 1)}_{EMM} |^{2},
\end{equation}
where $\Omega^{(\pm 1)}_{IMM}$ is the Rabi rate of the sampled intrinsic micromotion.  The amount of IMM sampled during the measurement is a function of the temperature of the ion.  For a single probe beam, oriented parallel to a single secular mode of motion with frequency $\omega_{r}$, the contribution to the Rabi rate ratio, to lowest order in the Lamb-Dicke parameter $\eta$ is given by \cite{Kellerthesis2015}
\begin{equation}
\label{eq: IMMonemode}
\frac{\Omega^{(\pm 1)}_{IMM}}{\Omega^{(0)}} = \frac{q}{4} \eta^{2} (2n +1) + \mathcal{O}(\eta^{4}),
\end{equation}
where $q = 2\sqrt2 (\omega_{r} / \Omega_{RF})$ is the Mathieu q-parameter and $n$ is the motional state of the mode.  For the case where we consider each of the six modes and each of the three beams, Eq.~(\ref{eq: IMMonemode}) is modified to include the fact that each beam will have a projection onto each mode of motion.  Taking the sum of the sampled IMM for each mode from each beam as the uncertainty we arrive at $\Delta \nu / \nu = 6 \times 10^{-21}$ when making measurements after ground state cooling.

\subsubsection{Influence of vertical EMM on the $\sigma$-port and $\pi$-port measurements}
Due to a phase imbalance between RF electrodes, the minimum EMM in the vertical direction is non-zero.  This vertical micromotion leads to a corresponding micromotion in the horizontal plane as follows.  Since the vertical micromotion is not zero at the minimum, we can think of the micromotion in the vertical direction leading to the ion sampling the horizontal RF field even when the average position is at the trap center.  This resulting micromotion is similar to the IMM associated with secular motion in that it changes phase every half cycle and occurs at twice the trap drive frequency and therefore is not detected by our measurements to first order.  A worst case estimate can be made by taking the amplitude of the vertical micromotion assuming the ion spends all of its time at the turning points of the motion.  This estimate is equivalent to the case in which the ion is displaced vertically from the RF minimum and a horizontal EMM is induced.  The vertical micromotion amplitude is given as $\beta / k_{L}$ where $\beta = 0.11$ is the average vertical EMM modulation index.  From the measured mode frequencies we can calculate the average radial confinement and the corresponding EMM for a given displacement from the RF minimum.

For a single $^{3}P_{1}$ beam (i.e. $\sigma$-port), the corresponding time-dilation shift is $\Delta \nu / \nu = -6.4 \times 10^{-20}$.  Taking the full shift as the uncertainty and adding the contribution from the $\pi$-port in quadrature we arrive at an uncertainty in the horizontal plane measurements due to the vertical micromotion of $\Delta \nu / \nu = 9.1 \times 10^{-20}$.

\subsubsection{EMM caused by ion displacement during the clock probe}
There can be additional micromotion during the clock pulse due to displacement of the ion away from the RF minimum caused by charging of the trap electrodes during initial cooling.  The magnitude of the displacement is approximately 2~nm over the typical 150 ms clock probe time.  Based on measurements of ion displacement during the clock probe discussed above, the 95 \% confidence bound on the excess micromotion time-dilation shift due to ion displacement during the clock probe is ~$\Delta \nu/ \nu = -2.5 \times 10^{-20}$.  Since we cannot detect the direction of the movement, this shift is added in quadrature as an additional uncertainty to the total EMM shift uncertainty.

\subsubsection{Summary of EMM uncertainty}
The total uncertainty on the EMM evaluation is shown in Table \ref{tab: EMMBudget}.  The EMM uncertainty is dominated by the unknown relative phase of the EMM along the three measurement directions.  This uncertainty can be reduced using a trap design with better control of the RF phase on each electrode and by using orthogonal $^{27}$Al$^{+}$ $^{3}P_{1}$ laser beams.
\begin{table}[h]
\caption{\label{tab: EMMBudget} Excess micromotion uncertainty budget.}
\begin{ruledtabular}
\begin{tabular}{lc}
Effect & Uncertainty $(\times 10^{-19})$ \\
\hline
Statistics, beam alignment, phase & 5.8 \\
Influence of vertical EMM & 0.9 \\
First-order Doppler EMM & 0.3 \\
Sampling of the intrinsic micromotion & 0.1 \\
\hline
Total EMM uncertainty & 5.9 \\
\end{tabular}
\end{ruledtabular}
\end{table}


\section{Secular Motion}
The time-dilation shift and uncertainty due to the secular motion of the ion have been evaluated following the procedure described in \cite{Chen2017PRL, Chen2017Thesis}.  After Doppler cooling, the probability $P(n)$ as a function of the motional Fock state is given by a thermal distribution
\begin{equation}
\label{eq:thermdist}
P_{th} (n) = \frac{1}{1+\bar{n}}\left(\frac{\bar{n}}{1+\bar{n}}\right)^{n} \,
\end{equation}
where $\bar{n}$ is the average occupation number.  For the case where $\eta \ll 1$ and for an infinitely long sideband cooling time, the resulting population distribution is also given by Eq.~(\ref{eq:thermdist}).  Under these conditions, the sideband thermometry technique is valid for determining the average occupation numbers in each motional mode after GSC \cite{Diedrich1989PRL, Monroe1995PRL}.

\begin{figure}[]
\includegraphics[width=1.0\columnwidth]{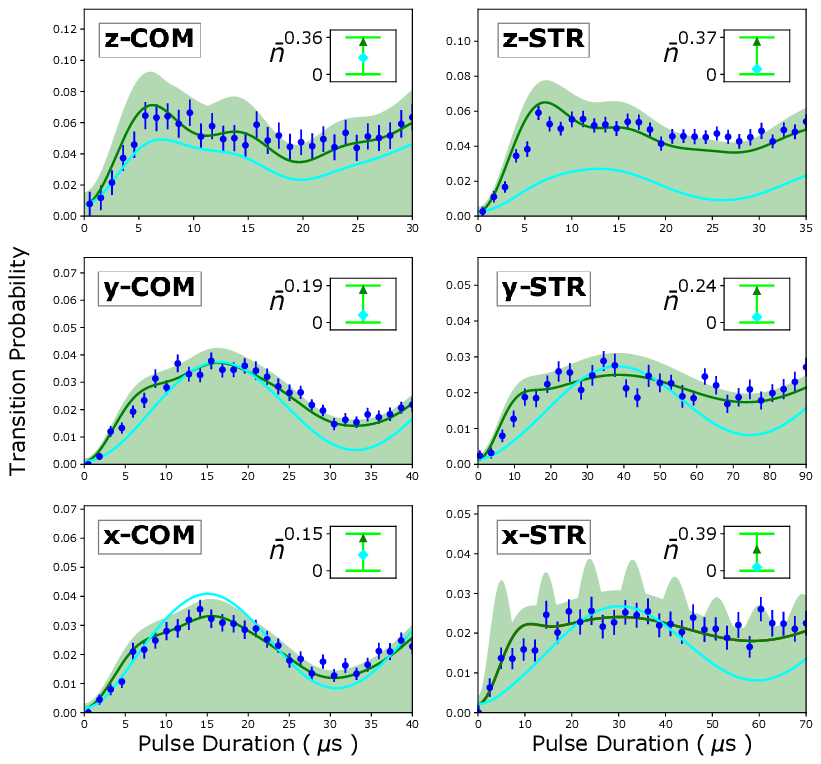}
\caption{\label{rsb_flopping}
Red sideband Rabi flopping on the $^{25}$Mg$^{+}$ $|^{2}S_{1/2}, F = 3, m_{F} = -3\rangle \leftrightarrow |^{2}S_{1/2}, F = 2, m_{F} = -2\rangle$ transition, after sideband cooling. Green, triple thermal distribution model; cyan, simulation.  The green shaded region indicates the residual Rabi flopping signals that are within the $95~\%$ confidence interval of the triple thermal distribution fit used to bound the time-dilation shift due to the secular motion.  The insets show the energy estimate from the simulation and fitting model, and the energy bounds that are used to calculate the clock uncertainty~\cite{Chen2017PRL,Chen2017Thesis}.}
\end{figure}

In the $^{27}$Al$^{+}$ clock, 342 sideband pulses are applied to the $^{25}$Mg$^{+}$ to cool all six secular modes of motion (57 pulses to each secular mode).  Due to the relatively large $\eta$ and finite number of cooling pulses, we find that the results from a simulation of the cooling process must be described using a more complicated distribution of motional states, as opposed to a single- or double-thermal distribution described elsewhere \cite{Diedrich1989PRL, Monroe1995PRL, Chen2017PRL}.  Here, for the purposes of establishing a conservative upper bound on the kinetic energy, we use a three-component thermal distribution, which is motivated by the results of the numerical simulation.  The resulting probability $P(n)$ is expressed as 
\begin{equation}
\label{triple_thermal_model}
\begin{array}{r@{}l}
P(n) &= \alpha P_{th}(n|\bar{n}_l) + \beta P_{th}(n|\bar{n}_m) + \gamma P_{th}(n|\bar{n}_h)\text{,}\\
\alpha &+ \beta + \gamma = 1\text{,} \\
\bar{n}_l &< \bar{n}_m < \bar{n}_h\text{,}
\end{array}
\end{equation}
where $P_{th}$ is given by Eq.~(\ref{eq:thermdist}) and $\bar{n}_l$, $\bar{n}_m$, and $\bar{n}_h$ denote average occupation numbers of three different thermal distributions.  To determine the secular motion kinetic energy requires more than sideband thermometry.  We fit red sideband (RSB) Rabi flopping data on each of the six motional modes to a model of the population obtained after 3D-GSC.  These results are shown in Fig. \ref{rsb_flopping} where the notation is the same as in \cite{Chen2017PRL}.

\begin{table*}[ht!]
\caption{\label{secular_modes}
Parameters of secular motion of a $^{25}\text{Mg}^+$-$^{27}\!\text{Al}^+$ two-ion pair during the clock operation. The trap axis is along $\hat{z}$ while $\hat{x}$ and $\hat{y}$ represent two mutually orthogonal transverse directions (see Fig.~\ref{fig:trapscheme}). In each axis, the COM mode is the mode in which the two ions move in the same direction while the STR mode is the mode in which the two ions move in the opposite directions. TDS, time-dilation shift; $\dot{\bar{n}}$, heating rate; $\bar{n}_0$, bounds of the average occupation number after 3D sideband cooling.
}
\begin{ruledtabular}
\begin{tabular}{ccccccc}
 & $\hat{x}$-COM & $\hat{x}$-STR & $\hat{y}$-COM & $\hat{y}$-STR & $\hat{z}$-COM & $\hat{z}$-STR \\ 
\hline
Frequency [MHz] & $3.31$ & $2.83$ & $3.97$ & $3.54$ & $1.47$ & $2.55$ \\
TDS/quantum [$10^{-18}$]\footnotemark[1] & $-0.114$ & $-0.518$ & $-0.092$ & $-0.521$ & $-0.066$ & $-0.098$ \\
$\dot{\bar{n}}$ [$s^{-1}$]\footnotemark[2] & $19.2(3.2)$ & $3.4(2)$ & $19.9(2.0)$ & $3.8(6)$ & $26.7(1.4)$ & $1.21(4)$\\
$\bar{n}_0$ [quantum]\footnotemark[3] & $0$ - $0.15$ & $0$ - $0.39$ & $0$ - $0.19$ & $0$ - $0.24$ & $0$ - $0.39$ & $0$ - $0.37$\\
Triple thermal fit reduced $\chi^2$ & $1.03$ & $1.19$ & $1.33$ & $1.72$ & $0.88$ & $3.27$\\
\end{tabular}
\end{ruledtabular}
\footnotetext[1]{including the shift due to the intrinsic micromotion in the transverse directions.}
\footnotetext[2]{$2\sigma$ statistical uncertainty.}
\footnotetext[3]{$95~\%$ confidence interval. The zero-point energy is not included.}
\end{table*}

To avoid over-fitting to the experimental data, we fix $\bar{n}_m$, $\bar{n}_h$, and the ratio $\beta/\gamma$, which are all derived from the simulation results.  In addition to the initial ion temperature after cooling, we have measured the motional heating rates for all six modes of the $^{25}$Mg$^{+}$ - $^{27}$Al$^{+}$ two-ion pair.  These results, along with the corresponding time-dilation shift are shown in Table~\ref{secular_modes}.  The 95\% confidence interval of the fit to $\bar{n}_{p,0}$ and the $2\sigma$ uncertainty of the $\dot{\bar{n}}_{p}$ measurements are used to assign the uncertainty in the time-dilation shift.  We assume a worst case scenario where the secular motion time-dilation shift in all of the motional modes is perfectly correlated when evaluating the uncertainty.

\section{Blackbody radiation}
\subsubsection{Stark effect}
The frequency shift due to blackbody radiation (BBR) arises from off-resonant coupling of the BBR to atomic levels via the Stark effect and the Zeeman effect~\cite{Gallagher1979PRL,Itano1981JP,Itano1982PRA_BBR}.  The frequency shift of an atomic level $|a\rangle$ due to an off-resonant monochromatic radiation source, $\mathcal{E}_0 \cos(\omega t)$, via the Stark effect is
\begin{equation}
\delta \nu_a = -\frac{1}{4 h}\mathcal{E}_0^2 \alpha_a(\omega)\text{,}
\end{equation}
where $h$ is the Planck constant, and $\alpha_a(\omega)$ is the scalar polarizability of the level $|a\rangle$ defined as
\begin{align}\label{polarizability}
\alpha_a (\omega) &= \frac{e^2}{m_e}\sum_j \frac{f_j}{\omega_j^2-\omega^2}\text{.}
\end{align}
Here $\omega_j$ and $f_j$ denote the frequency and the oscillator strength of transitions that connect to the energy level $|a\rangle$, respectively.
Therefore, the frequency shift of the transition $|a\rangle \rightarrow |b\rangle$ due to an off-resonant radiation field is given by
\begin{equation}
\label{transition_polarizability}
\Delta \nu = \delta\nu_b - \delta\nu_a = -\frac{1}{4 h}\mathcal{E}_0^2 \Delta\alpha_{a \rightarrow b}(\omega)\text{,}
\end{equation}
where $\Delta\alpha_{a \rightarrow b}(\omega)$ is the differential polarizability between two atomic states.
When an $^{27}\text{Al}^{+}$ ion is exposed to the BBR field from an environment at temperature $T$, the resulting shift of the clock transition due to the Stark effect can be calculated by integrating over the power spectrum of the BBR,
\begin{equation}\label{BBR_shift_Stark}
\Delta \nu^{(Stark)}_{clock} = \frac{-1}{4 \epsilon_0 \pi^3 c^3} \int_0^\infty \Delta \alpha_{clock}(\omega) \frac{\omega^3}{e^{\hbar \omega / k_B T}-1}d\omega\text{,}
\end{equation}
where $k_B$ and $\epsilon_0$ are the Boltzmann constant and vacuum permittivity, respectively.  The characterization of the BBR temperature environment that the ion experiences and the differential polarizability are both required to evaluate the frequency shift due to the BBR via the Stark effect.

\subsubsection{Zeeman effect}
The oscillating magnetic field from the BBR can also shift the clock transition frequency via the quadratic Zeeman effect.
The dominant source of the quadratic Zeeman shift is from the M1 transition $^{3}P_{0} \leftrightarrow ^{3}P_{1}$.  Although the transition frequency $\simeq 1.8$~THz is close to the mean BBR photon energy, the induced Zeeman shift due to BBR is small~\cite{Porsev2006PRA}.  The BBR frequency shift of the $^{27}\text{Al}^+$ clock transition related to the M1 transition $^{3}P_{0} \leftrightarrow ^{3}P_{1}$ is estimated to be less than $10\,\mu$Hz ($\approx 9 \times 10^{-21}$) and is negligible at the current clock systematic uncertainty~\cite{Safronova2011PRL}.
\subsection{Differential polarizability}
The differential polarizability was determined by measuring the frequency shift of the clock transition due to illumination of the ion with a $976$ nm fiber-pigtailed diode laser.  The parameters used to calculate the polarizability are summarized in Table~\ref{polarizability_measurement}.  The beam size at the position of the ion was characterized by measuring the frequency shift while translating the laser beam.  The beam profile is assumed to be gaussian and elliptical to calculate the intensity and the uncertainties in the beam waists listed in Table~\ref{polarizability_measurement} are derived from a fit to the beam profile.
\begin{table}[h]
\caption{\label{polarizability_measurement}Experimental parameters used to measure the differential polarizability.}
\begin{ruledtabular}
\begin{tabular}{cc}
Wavelength [nm] & $976$\\ 
Power [W] & $0.500(25)$ [$5.0~\%$]\\
Beam size $\omega_x$ [$\mu$m] & $109.8(4.8)$ [$4.4~\%$]\\
Beam size $\omega_y$ [$\mu$m] & $103.1(2.3)$ [$2.2~\%$]\\
Frequency shift [Hz] & $-65.8(1.5)$ [$2.3~\%$]\\
\end{tabular}
\end{ruledtabular}
\end{table}
For the evaluation of the uncertainty of the measured differential polarizability, we conservatively inflate the uncertainty of the beam area to $10~\%$ to account for possible deviations of the profile from gaussian.  Together with the $5~\%$ uncertainty of the laser power, this leads to $11.2~\%$ uncertainty in the intensity.  The mean-squared electric field of the laser is calculated to be $\langle \mathcal{E}_{976}^2 \rangle = 1.06(12) \times 10^{10}$ V$^2/$m$^2$.  Assuming the clock frequency shift is completely due to the AC Stark effect, the differential polarizability of the clock transition at $976$ nm is calculated using Eq. (\ref{transition_polarizability}),
\begin{align}
\Delta \alpha_{clock} (976\,nm) &= - \frac{2 h \Delta \nu}{\langle \mathcal{E}_{976}^2 \rangle} \\
& = 8.23(94)\times 10^{-42}\,\text{Jm}^2/\text{V}^2\text{.}
\end{align}
Following Eq. (5) in Ref.~\cite{Rosenband2006ProcEFTF}, we calculate the zero-frequency (DC) differential polarizability and its uncertainty using
\begin{widetext}
\begin{align}
\Delta\alpha_{clock}(\rm{DC}) &=  \frac{\Delta\alpha_{clock}(\lambda) - \frac{e^2}{m_e}\sum_i\frac{f_i}{\omega_i^2}(\epsilon_i + \delta_i^2/(1-\delta_i))}{1+\delta_0}\text{, and} \label{eq:diffpollam}\\
\sigma_{\Delta\alpha_{clock}(\rm{DC})} &= \left[ \left(\frac{\sigma_{\Delta \alpha_{clock} (\lambda)}}{1+\delta_0}\right)^2+\left(\frac{e^2/m_e}{1+\delta_0}\right)^2\sum_i\left(\frac{\sigma_{f_i} }{\omega_i^2}\left(\epsilon_i+\frac{\delta_i^2}{1-\delta_i}\right)\right)^2\right]^{0.5}\text{,}
\end{align}
\end{widetext}
where $\omega_i=2\pi c/\lambda_i$, $f_i$ and $\sigma_{f_i}$ are the transition frequency, oscillator strength, and uncertainty in the oscillator strength connecting either the $^{1}S_{0}$ or $^{3}P_{0}$ states to those given in Table II of Ref~\cite{Rosenband2006ProcEFTF}.  Equation (\ref{eq:diffpollam}) is exact in second-order perturbation theory, but it has been parameterized using the dimensionless quantities $\delta_0 = (\lambda_{0}/\lambda)^2$, $\delta_i = (\lambda_i/\lambda)^2$, and $\epsilon_i = (\delta_i-\delta_0)$ where $\lambda_{0}$ is a free parameter.  Substituting the measured value of $\Delta \alpha_{clock}(976 \rm{~nm})$ and choosing $\lambda_{0} = 174$~nm to minimize the uncertainty, we derive the differential polarizability of the clock transition at zero frequency to be $\Delta \alpha_{clock}(\rm{DC}) = (7.02 \pm 0.94) \times 10^{-42}$ Jm$^2$/V$^2$, which is in good agreement with the theoretical value of $\Delta \alpha_{clock}(\rm{DC}) = (8.2 \pm 0.8) \times 10^{-42}$ Jm$^2$/V$^2$ \cite{Safronova2011PRL}.  In a previous measurement of $\Delta \alpha_{clock}(\rm{DC})$ \cite{Rosenband2006ProcEFTF}, the ellipticity in the 1126~nm laser beam was not properly accounted for which led to a slight discrepancy between \cite{Rosenband2006ProcEFTF} and the result presented here.  The dynamic correction evaluated at 1126~nm is now in agreement with the theoretical value reported in Ref.~\cite{Safronova2011PRL}.  Given $\Delta \alpha_{clock}(\rm{DC})$, we estimate the differential polarizability for electromagnetic radiation at the wavelength $\lambda$ (when it is far-detuned from all transitions connecting to either clock states) to be
\begin{widetext}
\begin{align}
\Delta\alpha_{clock}(\lambda) &=  \Delta\alpha_{clock}(\rm{DC})(1+\delta_0) + \frac{e^2}{m_e}\sum_i\frac{f_i}{\omega_i^2}\left(\epsilon_i + \frac{{\delta_i}^2}{(1-\delta_i)}\right)\text{, and}\\
\sigma_{\Delta\alpha_{clock}(\lambda)} &= \left[ \left(\sigma_{\Delta \alpha_{clock}(\rm{DC})}(1+\delta_0)\right)^2+\left(\frac{e^2}{m_e}\right)^2\sum_i\left(\frac{\sigma_{f_i} }{\omega_i^2}\left(\epsilon_i+\frac{{\delta_i}^2}{1-\delta_i}\right)\right)^2\right]^{0.5}\text{.}
\end{align}
\end{widetext}



\subsection{Consideration of the quadratic Zeeman effect on the measurement of $\Delta \alpha_{clock}$}
We assume the quadratic Zeeman shift arrises from coupling to the M1 transition $^{3}P_{0} \leftrightarrow ^{3}P_{1}$ and only alters the energy of the $^{3}P_{0}$ state.  We estimate the magnitude of the quadratic Zeeman coefficient at $976$ nm 
\begin{equation}
\left |C^{967\,nm}_2 \right | \approx \frac{1.8}{308.2}\left |C^{dc}_2 \right | \approx 4.2 \times 10^{5}\,\text{Hz/Tesla}^2\text{,}
\end{equation}
and the resulting shift is $\approx 50$ mHz, which is within the uncertainty of the measured frequency shift caused by the 976 nm laser and does not change the result given in the previous section.

\subsection{Temperature}
The $^{27}$Al$^{+}$ ion is housed in a room temperature vacuum chamber.  During clock operation, the temperature of the trap wafer increases slightly due to resistive heating from trap RF currents.  To characterize the thermal environment experienced by the ion, three E-type thermocouples are installed inside the chamber; one is used to measure the temperature of the trap wafer and two others are used to measure the temperature of the post.  Another four sensors are used to measure the temperature of the vacuum chamber.  The highest temperature at $0.29$~W RF drive power (the nominal value used in the experiment), 295.65~K, was measured with a sensor mounted close to the RF resonator and is used as the upper bound of the temperature at the location of the ion, $T_{BBR}$.  The lower bound of $T_{BBR}$ is set by the ambient laboratory environment.  The lowest chamber temperature, 293.85~K, was observed at the position farthest away from the RF resonator and the magnetic field coils.  Temperatures are recorded once per minute during clock operation.  The thermocouple sensors have a specified accuracy of $1.7$~K and the temperature logger that converts the thermocouple voltage to a temperature has a specified accuracy of $0.5$~K.  These two effects contribute a $1.8$~K systematic uncertainty in the temperature measurements.  Combining all of these effects, the temperature at the location of the ion is bounded in a range $T_{BBR} = (292.05$ - $297.45)$~K.
\\
\subsection{BBR Summary}
Given the DC differential polarizability, the clock frequency shift due to BBR via the Stark effect can be obtained by integrating Eq.~(\ref{BBR_shift_Stark}), which gives
\begin{equation}
\Delta\nu^{(Stark)}_{clock} = \frac{-\pi\Delta\alpha^{(T)}_{clock}}{60 \epsilon_0 c^3}\left( \frac{k_B T}{\hbar}\right)^4\,\text{Hz,}
\end{equation}
where $\Delta \alpha^{(T)}_{clock}$ is the differential polarizability associated with a BBR field at the temperature $T$.  At $T \approx 300$~K, $\Delta\alpha^{(T)}_{clock} \simeq \Delta\alpha_{clock}(10~\mu \text{m}) = 7.03(94) \times 10^{-42}$ Jm$^2/$V$^2$.
From Eq.~(\ref{BBR_shift_Stark}), the uncertainty in the BBR shift is expressed as
\begin{widetext}
\begin{align}
\sigma_{\Delta\nu^{(Stark)}_{clock}} &= \left(\frac{\pi k_B^4}{60 \hbar^4 \epsilon_0 c^3}\right) \sqrt{\left( T^4\,\sigma_{\Delta\alpha^{(T)}_{clock}}\right)^2 + \left(4 \Delta\alpha^{(T)}_{clock}\,T^3 \Delta T\right)^2} \label{BBR_uncertainty_Stark}
\end{align}
\end{widetext}
where $\sigma_{\Delta \alpha^{(T)}_{clock}}$ and $\Delta T$ denote the uncertainties of the differential polarizability and the measured BBR temperature, respectively.  The first term in Eq. (\ref{BBR_uncertainty_Stark}) denotes the contribution to the total uncertainty due to the uncertainty in the differential polarizability, while the second term represents the contribution to the total uncertainty due to the uncertainty in the BBR temperature.  The fractional frequency shift and uncertainty due to BBR is $(\Delta \nu_{BBR} / \nu) = -(3.05 \pm 0.42) \times 10^{-18}$. 


\bibliography{suppreferences}